\documentclass[prl,superscriptaddress,twocolumn]{revtex4}
\usepackage{amssymb,amsfonts,amsmath}
\usepackage{epsfig,graphicx}\usepackage{mathrsfs}
\usepackage[pdftex,linkcolor=red]{hyperref}

\newcommand{\Tr}{\operatorname{Tr}}

\newcommand{\map}[1]{\mathscr{#1}}
\newcommand{\spc}[1]{\mathscr{#1}}

\newcommand{\Supp}{{\mathsf{Supp}}}
\newtheorem{theo}{Theorem}

\def\qed{$\blacksquare$ \newline}
\def\>{\rangle}
\def\<{\langle}
\def\map#1{\mathcal{#1}}
\def\Proof{{\bf Proof.~}}
\begin{document}

\title{Optimal design and quantum benchmarks for coherent state amplifiers}

\author{Giulio Chiribella}
\email{gchiribella@mail.tsinghua.edu.cn}
\author{Jinyu Xie}
\email{xiejy09@mails.tsinghua.edu.cn}

\affiliation{Center for Quantum Information, Institute for Interdisciplinary Information Sciences, Tsinghua University, Beijing 100084, China} \homepage{http://iiis.tsinghua.edu.cn}

\begin{abstract}
We establish the ultimate quantum limits to the amplification of an unknown  coherent state, both in the deterministic and probabilistic case, investigating the realistic scenario where the expected photon number is finite.
In addition, we provide the benchmark that  experimental realizations have to surpass in order to beat all classical amplification strategies and to demonstrate genuine quantum amplification.   Our result guarantees that a successful demonstration is in principle possible for every finite value of the expected photon number. 
\end{abstract} \maketitle

\maketitle

Continuous-variable quantum systems, such as coherent light pulses, are promising  information carriers  for the new quantum technology \cite{review1,review2}.  One of the cornerstones of continuous-variable quantum information is the  amplification of signals encoded into quantum states of the radiation field  \cite{micro2,micro3}.  Unlike classical amplifiers, quantum amplifiers are subject to fundamental  limits, typically expressed as  a reduction of the signal-to-noise ratio (SNR) as a function of the amplification parameter  \cite{linear,caves,cavesnew}.  Despite these limits,  quantum amplifiers are an essential piece of technology \cite{review3}, for they enable the detection of ultra-weak signals---such as gravitational waves---that would not trigger the detectors otherwise.   

Determining the ultimate quantum limits to amplification is both a topic of  immediate technological import and a  fundamental chapter of quantum theory, deeply connected with the no-cloning theorem, the uncertainty principle, and the quantum-classical transition in the limit of large amplification. Up to now, however, the performances of quantum amplifiers have been discussed mostly in classical terms (SNR), which are well suited for tasks such as signal detection, but less suited for applications in quantum information processing. 
 For example,  the role of the amplifier could be  to coherently copy quantum data  \cite{Andersen2005,telecloning} and to broadcast them to the users of a quantum internet \cite{kimble}.    For quantum tasks,  the most natural figure of merit is the fidelity between the desired output states and the states effectively produced by the amplifier, which can be interpreted operationally as the probability that the output state passes a test set up by a verifier who knows the input state.
  
In the fidelity setting, the works on optimal cloning of coherent states  \cite{Cerf2000-2,Lindblad2000,Cerf2000-3,Cochrane2004} give a first insight in the problem of optimal amplification, suggesting that two-mode squeezing should be  the best deterministic process allowed by quantum mechanics. 
  If confirmed in a realistic scenario, this conclusion would be  of high practical importance, as it would allow one to construct the best possible amplifiers using an optical element that is already in the toolbox of most laboratories.
 However, the optimality of two-mode squeezing, long conjectured, has never been proved without invoking strong simplifying assumptions, either on the nature of the amplifier---typically assumed to be Gaussian---or on the probability distribution of the states to be amplified---typically assumed to be uniform over all coherent states.    Both assumptions are far from trivial:  On the one hand, it is well known that non-Gaussian operations often outperform Gaussian ones, even for the manipulation of coherent states \cite{Cerf2005}. Hence, there is no \emph{a priori} reason to expect that the best amplifier of coherent states is Gaussian. From a fundamental point of view, any restriction on the allowed operations can hardly be satisfactory: if one wants to discover the ultimate quantum limits, one should not restrict the search to a subset, such as the subset of Gaussian operations, which has measure zero in the set of all possible operations.  On the other hand, assuming a uniform distribution over coherent states means assuming that the expected photon number is infinite, or equivalently, that there is no bound on the energy of the source producing the coherent pulses---a quite unphysical assumption.  In a realistic setting one can only have a large  photon number, and in order to know how large this number should be to be effectively treated as infinite, one needs to gain first a full grasp of the finite photon number scenario.
 
 Further motivation to go beyond the assumption of uniform distribution comes from the recent proposals of noiseless probabilistic amplifiers \cite{rl,marek,ampli1,ampli2,ampli3,ampli4}, whose performances are almost ideal for low photon numbers but  decay exponentially as the photon number increases.  In this case, it is most natural to test the performances of the amplifier on input states with low photon number, because these are the states where the amplifier is expected to work.  Furthermore, in order to claim the demonstration of a genuine quantum amplifier,  a real experiment should surpass the classical fidelity threshold (CFT) \cite{opt2,us,polzik,bagan}, i.e. the maximum fidelity achieved by ``classical" amplifiers that produce  an estimate of the input state and, conditional to the estimate,  reprepare amplified states. In the case of probabilistic amplifiers, where the photon number is necessarily finite, it would be unfair to compare the experimental fidelity with a lower CFT computed for the uniform distribution. However, despite the urge to have suitable criteria to assess the new experimental breakthroughs on probabilistic amplification \cite{ampli1,ampli2,ampli3,ampli4}, the correct value of the CFT for probabilistic quantum amplifiers has never been derived up to now.

In this  Letter we  establish the ultimate limits on the fidelity of quantum and classical amplifiers, treating both the deterministic and probabilistic case without making any assumption on the type of amplifying process, and without making the assumption of infinite expected photon number.  
We focus on the realistic scenario where the coherent states are distributed according to a Gaussian prior, which is the most studied case for applications in coherent-state quantum cryptography \cite{gross2002,Grosshans2003-2,ibli,luk,qi}, cloning \cite{Cochrane2004}, and teleportation or storage  \cite{opt2}.
 In the deterministic case, we show that the maximum quantum fidelity can be achieved through a two-mode squeezing process with the amount of squeezing depending critically on the variance of the prior. 
 In the probabilistic case, the critical behavior persists, with a dramatic effect: for variances  below the critical value the optimal amplifier becomes non-Gaussian and its fidelity can be arbitrarily close to 1. 
We  then provide the value of the classical fidelity treshold (CFT) that must be experimentally surpassed in order to demonstrate the implementation of a genuine quantum amplifier.  The value of the CFT is the same for both  deterministic and probabilistic protocols and, luckily, it guarantees that a successful demonstration  is possible for every finite value of the expected photon number. For example, for a gain $g=2$ and variance $1/3$, the value of the CFT is $50\%$, while the fidelity achieved by the optimal deterministic amplifier is $85\%$.  The general techniques developed in this work 
are not limited to quantum amplification, but apply more broadly to the optimization of quantum devices for any desired quantum task, including e.g. cloning, time reversal, and purification.  At this level, they establish a tight relation between the demonstration of genuine quantum processing and the advantage of entanglement  in the maximization of a suitable Bell-type correlation.

Let us start the derivation of our results.  We begin from a general problem:  finding the best physical process that approximates a desired transformation $\rho_x \mapsto  \psi_x $, where  $\{\rho_x\}_{x\in\mathsf X}$ is a set of (possibly mixed) input states, given with prior probabilites $\{p_x\}_{x\in\mathsf X}$, and  $\{\psi_x  =  |\psi_x\>\<\psi_x|\}_{x\in\mathsf X}$ is a set of pure target states.  Finding the best coherent-state amplifiers is a special case of this problem, corresponding to the input  $\rho_{\alpha}  =  |\alpha\>\<  \alpha|$ and the output $|\psi_\alpha\>  =  |g\alpha\>$, where  $g>1$ is the desired gain. 
To approximate the transformation $\rho_x\mapsto \psi_x$, we will consider the most general deterministic process, described by a quantum channel  (completely positive trace-preserving map) $\map C $. 
The performances of the channel will be ranked by  the average fidelity $  F  = \sum_{x\in\mathsf X}  p_x ~  \< \psi_x  |  \map C (\rho_x)  |\psi_x\>$.  In addition to the deterministic processes we will also consider probabilistic ones, described by quantum operations (completely positive trace non-increasing maps).   The average fidelity of a quantum operation $\map Q$, conditional on its occurrence, is given by $F'  =   \sum_{x\in\mathsf X}   p_x    \< \psi_x|   \map Q (\rho_x)  |\psi_x\>/ (\sum_{x'\in  \mathsf X}   p_{x'}  \Tr[ \map Q (\rho_{x'})])$.  
The optimal fidelity, defined as the supremum of the fidelity over all possible deterministic (probabilistic) processes, will be denoted by $F^{det}$  ($F^{prob}$).  

\begin{theo}[\cite{renner,supplemental,fiurasek}]\label{lem:upper} 
For deterministic processes, the optimal fidelity for the transformation  $\rho_x \mapsto \psi_x$ is given by
\begin{align}\label{upper}
 F^{det}  = &   \inf_{ \sigma >0,  \Tr[\sigma] =1 }   \|  A_{\sigma}\|_{\infty}   \\
\nonumber A_{\sigma}:  = &   \sum_{x\in\mathsf X}     ~     p_x  ~     |\psi_x\>\< \psi_x |   \otimes    ( \sigma^{-  \frac 12}   \rho_x  \sigma^{-\frac 12} )^T ,
\end{align} 
where $\|  A_{\sigma}\|_{\infty}$ denotes the \emph{operator norm} $ 
\|A_{\sigma}\|_{\infty}  :=   \sup_{  \|  \Psi \|  =  1 }   \< \Psi  | A_{\sigma} |\Psi\>$, and $T$ denotes the transpose.  
 
For probabilistic processes, the optimal fidelity is given 
\begin{align}\label{upperprob}
F^{prob}=  \|   A_\tau  \|_{\infty}     \qquad \tau  :=   \sum_{x\in\mathsf X}  p_x    \rho_x.     
\end{align}
\end{theo}
Theorem \ref{lem:upper} is a powerful tool for the optimization of quantum devices: since every quantum state $\sigma>0$ gives an upper bound on the fidelity, finding a channel that achieves any of these bounds means finding an optimal channel.

 In addition to the performances of the best quantum processes, it is important to know the CFT   for the transformation $\rho_x \to  \psi_x$.   The CFT is  the maximum  fidelity that can be achieved with a classical, measure-and-prepare protocol, where the input state is measured with a positive operator-valued measure (POVM) $ \{P_y\}_{y\in  \mathsf Y}  $ and, conditionally on  outcome $y$, a state $\rho'_y$ is prepared.  
  In the deterministic case, the fidelity of the protocol is the fidelity of the measure-and-prepare channel $\widetilde{\map C}  (\rho)  =\sum_{y\in \mathsf Y}    \Tr[  P_y  \rho]  ~ \rho_y' $.  In the probabilistic case, the POVM $\{  P_y\}_{y\in\mathsf Y}$    includes an outcome $y  =  ?$, conditionally to which no output state is produced.  The  fidelity is then the fidelity of the measure-and-prepare quantum operation $\widetilde Q (\rho)  =  \sum_{y  \in\mathsf Y, y \not = ? }   \Tr[  P_y  \rho]  ~ \rho_y'$.   In the following, the CFT will be denoted by $\widetilde F^{det}$ ($\widetilde F^{prob}$) in the deterministic (probabilistic) case.     
\begin{theo}\cite{supplemental}\label{lem:upperCFT} 
For deterministic protocols, the CFT for the transformation $\rho_x \to  \psi_x$ is given by
\begin{align}\label{upperCFT}
 \widetilde F^{det}  = &   \inf_{ \sigma >0,  \Tr[\sigma] =1 }   \|  A_{\sigma}\|_{\times}   
 \end{align} 
where  $\|   A_{\sigma} \|_\times$ denotes the \emph{injective cross norm} \cite{take} $\|A_{\sigma}\|_{\times}: =      \sup_{ \|  \varphi \|  =  \|  \psi\|  =    1}    \<  \varphi |  \< \psi  |  A_{\sigma}  |  \varphi\>  |\psi\>$. 

For probabilistic protocols, the CFT is given by  
\begin{align}\label{upperprobCFT}
\widetilde F^{prob}=  \|   A_\tau  \|_{\times}   .
\end{align}
\end{theo}

{\bf Remark: quantum-classical  gap  and Bell-type correlations.} Note that the trace of the separable operator $A_{\sigma}$  with a quantum state is a Bell-type correlation.  Remarkably,  Eqs. (\ref{upperprob}) and (\ref{upperprobCFT}) state that for probabilistic processes the gap between the quantum fidelity and the CFT  is equal to  the gap   between the maximum Bell correlation achievable with entangled states and the maximum Bell correlation achievable with separable states. This relation establishes a tight connection between the demonstration of genuine quantum processing and the violation of  suitable Bell-type inequalities.  

We are now ready to tackle the optimal design of quantum amplifiers and to find the corresponding CFT. To account for the prior information about the input, we introduce a  probability distribution $p(\alpha)$, normalized as $\int  \frac {{\rm d}^2\alpha} \pi   p(\alpha)  =1$.    The most popular choice for  $p(\alpha)$, typically considered in the literature \cite{gross2002,Grosshans2003-2,ibli,opt2,luk,qi},  is a Gaussian distribution  with  mean   $\alpha_0$  and variance   $V  =  1/\lambda$.    The idealized ``uniform prior" can be retrieved  here in the limit $\lambda \to 0$.   Note that it is not restrictive to consider  probability distributions centred around $\alpha_0  =  0$:  indeed, both in the deterministic and probabilistic case, the fidelity   does not change if one \emph{1)}  replaces the prior $p(\alpha)$ by $p(\alpha-  \alpha_0)$, \emph{2)} displaces the input state by $-\alpha_0$, and \emph{3)} displaces the output of the amplifier by $g \alpha_0$. 
 For $\alpha_0 = 0$, the Gaussian $p_{\lambda} (\alpha)  =  \lambda  e^{-\lambda|\alpha|^2}$ represents the distribution of coherent states generated by a classical oscillator obeying the Boltzmann distribution and $\<  n\>  = 1/\lambda$ is the expected photon number.  A controlled way to generate Gaussian-distributed  coherent states  is to prepare a two-mode squeezed state and perform a heterodyne measurement on one mode.

To determine the optimal deterministic amplifiers,
it is useful to assess first the performances that can be achieved using two-mode squeezing, i.e. using quantum channels of the form 
 \begin{align}\label{squeez}
\map C_r  ( \rho )  =  \Tr_{B}   [  e^{ r ( a^\dag b^\dag -   a b)  }   (  \rho \otimes |0\>\<  0|   )   e^{- r ( a^\dag b^\dag -   a b)  }  ]     ,
\end{align}  
where  $r$ is the squeezing parameter, $ a$ and $b$ are the annihilation operators of the input mode and of an ancillary mode, respectively, and $\Tr_{B}$   denotes the partial trace over the ancillary Hilbert space. Optimizing the value of the squeezing parameter one  obtains the fidelity \cite{supplemental} 
\begin{equation}\normalsize
F^{squeez}_{g,\lambda}=\left\{
\begin{aligned}&\ \frac{\lambda+1}{g^2},&&\lambda \le g -1 \\
& \frac{\lambda}{\lambda+(g-1)^2},&& \lambda > g-1 .\end{aligned}\right.   \label{lower}
\end{equation}
Note the discontinuity of the first derivative of the fidelity at the critical value $\lambda^{det}_c=  g-1$.   This value  separates two  different domains: for $\lambda  \le  \lambda^{det}_c$ the optimal amount of squeezing in Eq. (\ref{squeez})  is $r=   \cosh^{-1} \left(   \frac g{\lambda+1}\right) $, while for all values $\lambda > \lambda^{det}_c$ the  optimal value is $r=0$, corresponding to no squeezing at all.  In other words,   when the prior information about the input state is large (i.e. when the variance is small), the best amplifying strategy consists in leaving the state unamplified.     In the case of $1$-to-$2$ cloning, this fact was noted by Cochrane, Ralph, and Doli\'nska  \cite{Cochrane2004}, who assumed from the start cloning processes based on two-mode squeezing. 
Armed with Theorem \ref{lem:upper}, we are now in position to prove that no deterministic process can beat two-mode squeezing: 
\begin{theo}{\bf (Optimal design of deterministic amplifiers \cite{supplemental})}\label{theo:opt}
 Two-mode squeezing is the  best deterministic process for the amplification of  Gaussian-distributed coherent states.  
\end{theo}  

For probabilistic amplifiers, however, the situation is very different.  Evaluating Eq. (\ref{upperprob}) we get \cite{supplemental}
\begin{equation}
F^{prob}_{g,\lambda} =\left\{
\begin{aligned}&\ \frac{\lambda+1}{g^2},&&\lambda \le g^2 -1 \\
& ~1 && \lambda > g^2-1 .\end{aligned}\right.   \label{probabilistic}
\end{equation}
The difference with the deterministic case is dramatic: above the the critical value $\lambda_c^{prob}  =  g^2-1$  probabilistic processes allow for noiseless  amplification.   Fidelity arbitrarily close to $F^{prob}_{g,\lambda}$ can be reached as follows: 
\begin{theo}{\bf (Optimal design of probabilistic amplifiers \cite{supplemental})}\label{theo:probopt}
 The  best probabilistic amplifier for Gaussian-distributed coherent states is 
 \begin{enumerate}
 \item for $\lambda \le  \lambda^{det}_c$, the two-mode squeezer (\ref{squeez}) with squeezing parameter $  r =  \cosh^{-1} [g/(\lambda+1)] $
 \item   for $  \lambda^{det}_c<\lambda \le  \lambda_c^{prob}$, a quantum operation $\map Q_N(\rho)  = Q_N \rho Q_N$ with $   Q_N  \propto     \sum_{n=0}^N   [(\lambda + 1)/g]^{n}   |n\>\<  n|$, achieving  fidelity $F^{prob}_{g,\lambda}  =  (1+\lambda)/g^2$  exponentially fast in the limit $N \to \infty$ 
 \item  for $\lambda > \lambda_c^{prob}$,  a quantum operation $\map Q_N(\rho)  = Q_N \rho Q_N$ with $   Q_N  \propto     \sum_{n=0}^N   g^{n}   |n\>\<  n|$, achieving the fidelity $F_{g,\lambda}^{prob}  = 1$ exponentially fast in the limit $N \to \infty$. 
  \end{enumerate}  
\end{theo}  
Note that for $\lambda >  g-1$ the optimal quantum operations are non-Gaussian, whereas for $\lambda = 0$  (``uniform prior")  the optimal deterministic and probabilistic amplifiers coincide and are Gaussian. Noiseless amplification is only possible when  the expected photon number is finite.

Suppose now that an experiment aims at demonstrating quantum amplification---or equivalently, cloning---of a coherent state. Thanks to  Theorem \ref{lem:upperCFT}, we can easily find the analytical expression of the CFT, also specifying the best measure-and-prepare channel.  The result applies to both deterministic and probabilistic protocols, and,  as an extra bonus,  provides a coincise  derivation of the quantum benchmark for teleportation and storage of coherent states found by Hammerer, Wolf, Polzik, and Cirac \cite{opt2}, which is retrieved here in the special case  of no amplification ($g=1$).  
   
\begin{theo}[Benchmark for quantum amplifiers \cite{supplemental}]\label{theo:bench}
The CFT for the  amplification of Gaussian-distributed coherent states is given by  
\begin{equation}\label{benchmark}
\widetilde F_{g,\lambda}  
=\frac{1+\lambda}{1+\lambda+g^2}
\end{equation}
both for deterministic and probabilistic protocols.  The above value is achieved by a heterodyne measurement $P(\hat \alpha)  \frac{d^2 \hat \alpha }\pi =  |\hat \alpha\>\<  \hat \alpha |  \frac{d^2 \hat \alpha }\pi$  followed by the conditional preparation of the coherent state  $ \left|  \frac{   g \hat \alpha}{1+\lambda }   \right\>$.   
\end{theo}

 Eqs. \ref{lower}, \ref{probabilistic} and \ref{benchmark} represent  good news for experimental demonstrations: they prove that genuine quantum amplification can be demonstrated \emph{for every finite value of the expected photon number}.  As an illustration, consider the demonstration  of probabilistic amplification provided by Zavatta, Fiur\'a\v cek and Bellini in Ref. \cite{ampli4}.        
  In this case, the amplifier is designed to achieve gain $g=2$. By Eq. (\ref{probabilistic}),   noiseless amplification requires at least $\lambda  \ge 3$, which is actually a reasonable value in the experiment  (choosing $\lambda = 3$  puts the maximum amplitude tested in the experiment, $|\alpha_{\max}|^2  \approx 1.0$, at three standard deviations from the mean photon number $\< n \>  =  1/3$, effectively cutting off the values $|\alpha|  >1$).    For $\lambda = 3$, Eqs. (\ref{lower})  and (\ref{probabilistic})  give $F^{squeez}_{g=2,\lambda=3}  =  85\%$ and $\widetilde F_{g=2,\lambda=3}  = 50\%$ for the fidelity of the best deterministic amplifier and for the CFT, respectively \cite{nota}.  The average of the experimental fidelities $F_{exp}  \approx 0.99/0.91/0.67$, corresponding to the amplitudes $|\alpha|\approx 0.4/0.7/1.0$, gives a value that is well above the benchmark for genuine quantum processing, but also very close to the value that can be achieved by deterministic amplifiers.  
One should observe, however, that the small number of values of $|\alpha|$ probed in the experiment precludes an accurate data analysis, as the average over few values of $\alpha$ is very sensitive to statistical fluctuations. 
Our analysis suggest that, although the available data show a neat quantum advantage over measure-and-prepare strategies, further experimental investigations would be desirable to enable a statistically significant analysis of the advantage of probabilistic amplifiers.    
To guarantee a fair sampling, the ideal setup would be to test the amplifier on Gaussian-distributed coherent states generated randomly by a heterodyne measurement on one side of a two-mode squeezed state.

{\bf The classical limit of quantum amplifiers.}  For $\lambda \le g-1$, the gap between the quantum fidelity and the CFT is equal to the gap between entangled and separable states in the Bell correlation $\<A_{\tau}\>$.  The gap vanishes in the limit $g \to \infty$,  and  the fundamental reason  is that an amplifier with infinite gain is classical, like a cloning device producing infinite clones \cite{bae,me,defi}.  
  This point is made very clear by our results:  denoting by $\map C_{g,\lambda}$ and by $\widetilde {\map C}_{g,\lambda}$ the optimal quantum amplifier and the optimal measure-and-prepare amplifier,  for $\lambda  \le  g- 1$ we have the remarkable relation \cite{supplemental}
    $ 
  \widetilde {\map C}_{g,\lambda}      =     \map A_{\frac g{\sqrt {g^2  +  (\lambda + 1)^2}}  } {\map C}_{\sqrt {g^2  +  (\lambda + 1)^2},\lambda},
$  
where $\map A_\eta$  is the attenuation channel transforming the coherent state $ |\alpha\>$  into $|\eta  \alpha\>$, $\eta \le 1$.  In words, the best measure-and-prepare  strategy with gain $g$ is equivalent to the best quantum strategy with gain $g' = \sqrt{g^2  +  (\lambda + 1)^2}$, followed by an attenuation of $\eta  =  g/ \sqrt {g^2  +  (\lambda + 1)^2}  $ that reduces the gain from $g'$ to $g$.  When the desired gain is large compared to the prior information available ($g \gg \lambda $) we have $g'  \approx g$ and $\eta \approx 1$, which imply $\widetilde {\map C}_{g,\lambda}  \approx  \map C_{g,\lambda}$. 


In conclusion, we established the ultimate quantum limits to the deterministic and probabilistic amplification of Gaussian-distributed coherent states, without making any assumption on the nature of the amplifier and without making the unrealistic assumption of uniform distribution over coherent states. For probabilistic amplifiers, we discovered the presence of a critical value of the expected photon number, below which noiseless amplification becomes possible.  Furthermore, we provided the quantum benchmark that has to be surpassed in order to establish the successful  experimental demonstration of a  genuine quantum amplifier.  Our results show an intriguing link between  genuine quantum amplification and  the maximization of a suitable Bell-type correlation, and,  in addition, they guarantee that a successful demonstration is possible for any finite value of  the expected photon number.  


\emph{Acknowledgments.}   This work is supported by the National Basic Research Program of China (973) 2011CBA00300 (2011CBA00301),  by the 1000 Youth Fellowship Program of China, and by the National Natural Science Foundation of China through Grants 61033001 and 61061130540.  We acknowledge the support of Perimeter Institute for Theoretical Physics, where this work was started.  Research at Perimeter Institute for Theoretical Physics is supported in part by the Government of Canada through NSERC and by the Province of Ontario through MRI. We thank the anonymous referees for inspiring a significant strengthening of our results, G Adesso and M Bellini for their advise on the comparison with experimental works and S Pirandola for comments on an earlier version of the manuscript.

\appendix  

\begin{widetext}

\section{Proof of Theorem 1:  General expression for the optimal quantum fidelity}\label{sec:lemmauno}
\begin{enumerate}
\item \emph{Deterministic case.}  
For a generic quantum channel $\map C$ and for an arbitrary quantum state $\sigma >0$,  it is easy to prove the upper bound 
\begin{align}\label{upper}
\nonumber F  \le &    \|  A_{\sigma}\|_{\infty}   \\
 A_{\sigma}:  = &  \sum_{x\in \mathsf X}  p_x  ~         |\psi_x\>\< \psi_x |   \otimes    \left( \sigma^{-  \frac 12}  \rho_i       \sigma^{-\frac 12}\right)^T .
\end{align} 
The proof runs as follows:  
\begin{align*}
F & =     \sum_{x\in \mathsf X}  p_x  ~       \<  \psi_x |  \map C  \left[ \left(    \sigma^{\frac 12} \right)   \left ( \sigma^{-\frac 12}   \rho_x   \sigma^{-\frac 12}  \right )    \left(  \sigma^{\frac 12}\right)  \right] |\psi_x\>  \\
 & =       \sum_{x\in \mathsf X}  p_x  ~         ~   \Tr\left [    |\psi_x    \>\<  \psi_x |  \otimes     \left(\sigma^{-\frac 12}  \rho_x  \sigma^{-\frac 12}     \right)^T  \Phi_{\sigma,\map C}  \right] \\
 &  =  \Tr[A_{\sigma} \Sigma_{\map C}],
 \end{align*}
 where $\Phi_{\sigma,\map C} $ is the quantum state defined by 
 \begin{align*}
 \Phi_{\sigma,\map C}  :  =    (\map C \otimes \map I)    (   |\sigma^{\frac 12}\>\!\>\<\!\< \sigma^{\frac 12}|)  \qquad |\sigma^{\frac 12}\>\!\>:  =  \sum_{m,n}   \< m|  \sigma^{\frac 12}   |n\> ~  |m\>|n\>.                
 \end{align*}
The bound of Eq. (\ref{upper}) then follows from the inequality $|\Tr [A_{\sigma} \Phi_{\sigma,\map C}] |  \le  \|  A_{\sigma} \|_{\infty}$, valid for every quantum state $\Phi_{\sigma,\map C}$.  Hence, we conclude that the maximum of the fidelity over all quantum channels, denoted by $F^{det}$  satisfies 
\begin{align}\label{upperinf}
F^{det}  \le  \inf_{\sigma>  0,\Tr[\sigma]=1}  \|  A_{\sigma} \|_{\infty}.
\end{align}
On the other hand, using the duality of semidefinite programming, one can show that the bound can be achieved.  In the case where the input and ouput states live in finite-dimensional Hilbert spaces, the proof was given  by K\"onig, Renner, and Schaffner in Ref.\cite{renner}. For completeness, we present it here in the language of our paper.  
Without loss of generality, we assume that the average input state 
\begin{align*}
\tau  :=  \sum_{x\in\mathsf X}  p_x  ~  \rho_x
\end{align*} 
is strictly positive [the latter condition can be imposed  by restricting the action of the channel $\map C$ to the support of $\tau$].   In this case, the fidelity can be written as 
\begin{align}\label{boh}
F  =  \Tr[ A  C ]   \qquad A:  =  \sum_{x\in\mathsf X}  p_x  ~  |\psi_x\>\< \psi_x |   \otimes   \rho^T_x,\quad     C : =  ( \map C   \otimes \map I) ( |I_{in}\>\!\>\<\!\< I_{in} | ), \quad | I_{in}\>\!\>  : =\sum_{n  = 1}^{\dim (\spc H_{in})}   |n\>|n\>.   
\end{align} 
where $C$, the Choi operator of the channel $\map C$, satisfies the normalization condition   
\begin{align}\label{norm}
\Tr_{out}  [C]  =  I_{in},
\end{align} 
$\Tr_{out}$ and $I_{in}$ denoting the partial trace on the output Hilbert space and the identity on the input Hilbert space, respectively. 
Since every positive operator $C\ge0$ satisfying Eq. (\ref{norm}) is the Choi operator of some channel,  the maximum fidelity can be computed by the semidefinite program 
\begin{align*}
F^{det}  = \max_{  C  \ge 0  ,  \Tr_{out} [ C]  = I_{in}}   \Tr[  C A] ,
\end{align*}
whose value, by strong duality, is equal to 
\begin{align*}
F^{det} &  = \min_{  \Lambda  \ge 0 , I_{out}  \otimes \Lambda  \ge  A}    \Tr[\Lambda].
\end{align*}
Note that, actually, $\Lambda$  must be strictly positive, because the average state $\tau$ is strictly positive.    
Defining $  \Lambda  =  t   \sigma^T$, where $t= \Tr[\Lambda]$ and $\sigma>0$ is a density matrix, we then have  
\begin{align}
\nonumber F^{det}   &=    \min \left\{t\ge 0 ~|~  \exists  \sigma > 0, \Tr[\sigma] = 1,   t  (I_{out} \otimes I_{in}) \ge  \left(  I_{out} \otimes \sigma^{-\frac 12}  \right)^T   A  \left(I_{out} \otimes \sigma^{-\frac 12}  \right)^T  \right\}   \\ 
\nonumber &=    \min_{\sigma > 0, \Tr[\sigma] = 1}  \left\|   \left(I_{out} \otimes \sigma^{-\frac 12}  \right)^T   A  \left(I_{out} \otimes \sigma^{-\frac 12}  \right)^T  \right\|_{\infty}\\
 &=    \min_{\sigma > 0, \Tr[\sigma] = 1}  \left\|    A_\sigma   \right\|_{\infty}\label{dua}
 \end{align}
Hence,  in finite dimensions we have a guarantee that the bound  of Eq. (\ref{upperinf}) can be achieved by a quantum channel.   When the input and output Hilbert spaces are infinite dimensional, we show that, in most relevant cases, one can reduce the problem to the finite dimensional case by truncating the dimension.   The technical details of the truncation are provided in the last section of this supplemental material. 
\item {\em Probabilistic case \cite{fiurasek}.}   For an arbitrary quantum operation $\map Q$, the fidelity is given by  
\begin{align*}
F     &=    \frac {  \sum_{x\in\mathsf X}      p_x~         \<   \psi_x  |   \map Q   (\rho_x)  |\psi_x\> }  { \Tr [  \map Q (\tau)]}.
\end{align*}

Following the same proof  as in the deterministic case, we get  
\begin{align}\label{ausiliaria}
F     &=   \frac{\Tr [  \Phi_{\tau,\map Q}    A_{\tau} ]}  { \Tr [  \map Q (\tau)]}  =  \frac{\Tr [  \Phi_{\tau,\map Q}    A_{\tau} ]}  { \Tr [  \Phi_{\tau, \map Q}]}  =  \Tr[ \Sigma_{\tau,\map Q}   A_{\tau} ] \qquad  \Sigma_{\tau,\map Q}  : =  \Phi_{\tau,\map Q}/ \Tr[\Phi_{\tau,\map Q}].
\end{align}
  Hence, we have the upper bound  
 $F    \le  \|       A_{\tau}  \|_\infty$.    
In finite dimensions, the bound can be achieved by taking the eigenvector of $A_\tau$ with maximum eigenvalue, denoted by $  |\Psi\>$, and using the state 
\begin{align*}
|\widetilde \Psi  \>   :  =  \frac {  (  I_{out}\otimes  \tau^{-\frac 12})^T  |\Psi\>}{\|   (  I_{out}\otimes  \tau^{-\frac 12})^T  |\Psi\>\|} 
\end{align*}  
 as the resource state in a probabilistic teleportation protocol. In infinite dimensions, the optimal fidelity is achieved in the limit, using approximate teleportation.  Note that $A_{\tau}$ itself may have only approximate eigenvectors.   
\end{enumerate}
 \qed

\section{Proof of theorem 2: general expression for  the CFT}

\begin{enumerate}
\item {\em Deterministic case.}  For a generic measure-and-prepare channel $\widetilde{\map C}$ and for every state $\sigma  > 1$,  it is easy to prove the upper bound 
\begin{align}\label{uppermp}
\nonumber \widetilde F  \le &    \|  A_{\sigma}\|_{\times}   \\
 A_{\sigma}:  = &    \sum_{x \in\mathsf X}    p_x    ~      |\psi_x\>\<  \psi_x |   \otimes    \left(  \sigma^{-  \frac 12}    \rho_x      \sigma^{-\frac 12} \right)^T.
\end{align} 

The proof is the same as the proof of  Lemma  1, with the only difference that now the state $\Phi_{\sigma,\widetilde{\map C}}  =   (\widetilde{\map C} \otimes \map I)  (  |\sigma^{\frac 12}\>\!\>\<\!\<  \sigma^{\frac 12}|)$ is separable.   By definition of the injective cross norm, we have  $\Tr[A_{\sigma}  \Phi_{\sigma, \widetilde{\map C} } ]  \le  \|A_{\sigma}\|_\times$, for every separable quantum state $\Phi_{\sigma,\widetilde{\map C}}$.   Hence, the CFT will be bounded as 
\begin{align}\widetilde F^{det}   \le  \inf_{\sigma>  0, \Tr[\sigma] =1}    \|  A_{\sigma}  \|_{\times}.
\end{align}  

Like in the proof of Lemma 1, we can use the duality of semidefinite programming to show that the upper bound is actually an equality.  Again,  we first consider first the case where the input and output Hilbert spaces are finite-dimensional Hilbert spaces, and the average input state 
\begin{align*}
\tau  :=  \sum_{x\in\mathsf X}  p_x  ~  \rho_x
\end{align*} 
is strictly positive.  
In this case, the fidelity can be written as 
\begin{align}\label{boh}
F  =  \Tr[ A  \widetilde C ]   \qquad A:  =  \sum_{x\in\mathsf X}  p_x  ~  |\psi_x\>\< \psi_x |   \otimes   \rho^T_x,\quad    \widetilde C : =  ( \widetilde {\map C}   \otimes \map I) ( |I_{in}\>\!\>\<\!\< I_{in} | ),
\end{align} 
where the Choi operator $\widetilde C$ is now separable. To turn the separability condition into a semidefinite program, we now use the $n$-extendability condition of Ref. \cite{parrillo}, stating that $\widetilde C$ is separable if and only if, for every $n \in\mathbb N$ there exists an $n$-symmetric extension, that is, an operator $\widetilde C_n $  on  $\spc H_{out}^{\otimes N}  \otimes \spc H_{in}$ such that 
\begin{enumerate}
\item   $\widetilde C_n$ extends $\widetilde C$, i.e.  
\begin{align*}
\Tr_{n-1} [\widetilde C_n]  =  \widetilde C
\end{align*}  
($\Tr_{k} $ denoting the partial trace over the first $k$ output spaces), and 
\item $\widetilde C_n$ is invariant under permutation of the outputs. 
\end{enumerate}  
Permutation invariance can be expressed as    $(\Pi_n \otimes \map I_{in}) (\widetilde C_n)  =  C_n$, where $\Pi_n$ is the permutation-twirling
\begin{align*}
\Pi_n (\rho)  =  \frac 1 {n ! } \sum_{\pi\in  S_n}    U_\pi   \rho    U_{\pi}^\dag,
\end{align*}            
$U_{\pi}$ being the unitary operator that implements the permutation $\pi \in S_n$ of the $n$  output spaces.
 We can then express the maximum fidelity over all measure-and-prepare channels as  
\begin{align*}  
\widetilde F^{det}  &  =  \inf_{n\in\mathbb N} \quad   \max_{   \widetilde C_n  \ge 0  , (  \Pi \otimes \map I_{in} )  (\widetilde C_n)  =  \widetilde C_n,   \Tr_{n}[\widetilde C_n]   = I_{in}}   \Tr [   \widetilde C_n   (I_{n-1}  \otimes A) ]  \\
&  =  \inf_{n\in\mathbb N} \quad   \max_{   C_n  \ge 0  ,    \Tr_{n}[ C_n]   = I_{in}}   \Tr [   C_n     (\Pi_n \otimes \map I_{in})(I_{n-1}  \otimes A) ]      
\end{align*}
Using strong duality for the maximization over $C_n$, we obtain  
\begin{align*}  
\widetilde F^{det}  &  =  \inf_{n\in\mathbb N} \quad   \min_{ \Lambda_n  \ge 0  ,    I_n \otimes \Lambda_n   \ge    (\Pi_n \otimes \map I_{in})  (I_{n-1}\otimes A)  }   \Tr [  \Lambda_n ] . 
\end{align*}
Now, since $\tau $ is strictly positive, also $\Lambda_n$ must be strictly positive. Writing $\Lambda_n  =  t_n   \sigma $, with $t_n  =  \Tr [\Lambda_n]$ and $\sigma>0$,   we have 
\begin{align*}  
\widetilde F^{det}  &  =  \inf_{n\in\mathbb N}\quad \min \{    t_n ~|~    \exists  \sigma  >  0,     \Tr[\sigma ]=1,   t(  I_n \otimes  I_{in})    \ge     (\Pi_n \otimes \map I_{in})  (I_{n-1}\otimes A_\sigma)      \}\\
&  =    \inf_{n\in\mathbb N}    \quad    \min_{\sigma  >  0  , \Tr[\sigma]=1  }    \|    (\Pi_n \otimes \map I_{in} )  ( I_{n-1} \otimes A_\sigma)    \|_{\infty}  \\
&   =    \inf_{n\in\mathbb N}    \quad    \min_{\sigma  >  0  , \Tr[\sigma]=1  }   ~  \max_{\rho_n  \ge 0 , \Tr[\rho_n]=1 }     \Tr \{  \rho_n  ~ [ (\Pi_n \otimes \map I_{in} )  ( I_{n-1} \otimes A_\sigma)  ]\} \\  
&   =\min_{\sigma  >  0  , \Tr[\sigma]=1  }    \quad     \inf_{n\in\mathbb N}        ~  \max_{\rho_n  \ge 0 ,    \Tr[\rho_n]=1,  (\Pi_n \otimes \map I_{in} )   (\rho_n)   =  \rho_n }     \Tr \{    \Tr_{n-1}  [\rho_n] ~  A_\sigma  \}  \\  
&  =  \min_{\sigma  >  0  , \Tr[\sigma]=1  }    \quad \max_{\rho ~{\rm separable} }   \Tr [\rho  A_{\sigma}]\\
&  \equiv  \|  A_\sigma  \|_{\times}.
\end{align*}

The validity of the formula $\widetilde F^{det}  =  \inf_{\sigma >0, \Tr[\sigma]  =1}\|  A_\sigma \|_\times $ in the case where the input and output Hilbert spaces are infinite dimensional can be proved using the truncation argument provided in the end of this supplementary material.
\item {\em Probabilistic case.}   
 Inserting a measure-and-prepare quantum operation $\widetilde Q$ into Eq. (\ref{ausiliaria}), we get  $ \widetilde F    =   \Tr[  A_\tau  \Sigma_{\tau,\widetilde {\map Q}} ]$, where $\Sigma_{\tau,\widetilde {\map Q}}$ is a separable state.   This  implies the bound  $\widetilde F^{prob} \le \|  A_\tau\|_{\times}  $.    
 In finite dimensions, a quantum operation that achieves the bound can be obtained by taking  two unit vectors  $ |\psi\> \in \spc H_{out}$ and $|\bar \varphi\> \in\spc H_{in}$ such that $\| A_{\tau}\|_{\times}  =  \<  \psi|  \<  \varphi|  A_{\tau}  |\psi\>  |\varphi\>$, and by defining $\widetilde {\map Q}  (\rho)   \propto    |\psi\>\<  \psi|    \<   \varphi|  \tau^{-\frac 12}  \rho  \tau^{-\frac 12}  |\varphi\> $.  In infinite dimensions, one may have to truncate the vector $\tau^{-\frac 12}  |\varphi\>$ to a finite dimensional subspace to make it normalizable.  Letting the dimension of the subspace grow, one obtains a sequence of quantum operations with fidelity converging to $\widetilde F^{prob}$.  
\end{enumerate}
 \qed

\section{Proof of Eq. (6): The performances of two-mode squeezing}  

A parametric amplifier $\map C_r  ( \rho )  :=  \Tr_{B}   [  e^{ r ( a^\dag b^\dag -   a b)  }   (  \rho \otimes |0\>\<  0|   )   e^{- r ( a^\dag b^\dag -   a b)  }  ] $  satisfies the covariance property    
\begin{align*} 
\map C_r ( D(\alpha)  \rho   D(\alpha)^\dag)  =  D( \alpha \cosh r )  \map C_r  (   \rho )  D^\dag(\alpha\cosh r) \qquad  \forall \alpha\in\mathbb C,
\end{align*}
for every trace-class operator $\rho\in\map T(\spc H)$.  
Moreover, we have 
\begin{align}\label{param} 
\map C_r  (  |0\>\< 0|)   =  (1-x)\sum_{n=0}^{\infty}  x^n  |n\>\<  n  |   :  =  \rho_x \qquad x  =  \tanh^2 r.
\end{align}  
Combining these two facts, the amplification fidelity of the channel $\map C_r$ is given by  
\begin{align*}
F^r_{g,\lambda} & =         \int_{\alpha \in  \mathbb C}  ~   \frac {{\rm d}^2  \alpha }\pi   ~   \lambda e^{-\lambda  |  \alpha|^2}   ~  \<  (g  -\cosh r)   \alpha  |   \rho_x  |   (g  -\cosh r)   \alpha \> \\
& =         \int_{\alpha \in  \mathbb C}  ~   \frac {{\rm d}^2  \alpha }\pi   ~  \frac{ \lambda}{  \cosh^2 r}   e^{-\lambda  |  \alpha|^2}   ~e^{-  \frac{(  g  - \cosh r )^2  |\alpha|^2 }{\cosh^2 r}  }   \\
&  = \frac {\lambda}{   \lambda  \cosh^2 r +  (  g  - \cosh r )^2}.    
\end{align*} 
The maximum of the function $F^r_{g,\lambda}$  is achieved by $  \cosh r =   g/(\lambda +  1) $ when $  g  \ge  \lambda + 1 $ and by $\cosh r  = 1$  otherwise, thus giving  
\begin{equation}\label{twomode}
F^{opt}_{g,\lambda}=\left\{
\begin{aligned}&\ \frac{\lambda+1}{g^2},&&\lambda \le g -1 \\
& \frac{\lambda}{\lambda+(g-1)^2},&& \lambda > g-1 .\end{aligned}\right.  
\end{equation}
which is what we wanted to prove.

\section{Proof of Theorem 3: Optimal design of deterministic amplifiers}

\Proof  We show that the performances of two-mode squeezing, given by Eq. (\ref{twomode}), are the best among the performances of all quantum channels.   
To this purpose, our strategy is  to find a state $\sigma$ such that  the upper bound 
provided by Eq. (\ref{upper}), matches the lower bound of Eq. (\ref{twomode}).

 As an ansatz, we  assume  $\sigma$ to be a thermal state, of the form  
\begin{align}\label{sigmax}
\sigma_x :=  (1-x)  \sum_{n=0}^{\infty}  x^n  |n\>\<  n| ,
\end{align} 
so that the operator $A_{g,\lambda,\sigma}$ becomes 
\begin{align}
A_{g,\lambda,x}  :=  \frac{\lambda}{1-x}\int\frac{d^2\alpha}{\pi} {}e^{-(\lambda+1-\frac{1}{x})|\alpha|^2}|g\alpha\rangle\langle g\alpha|
\otimes\left|\frac{\bar \alpha}{\sqrt{x}}\right\rangle\left\langle\frac{\bar \alpha}{\sqrt{x}}\right|  .
\end{align}      
 The  operator norm of $A_{g,\lambda,x}$ can be computed using the relation $\| A_{g,\lambda,x} \|_{\infty}  =  \lim_{p\to\infty} \left(  \Tr  |A_{g,\lambda,x}|^p\right)^{\frac 1p}$.  For each fixed $p$, the calculation consists only of  Gaussian integrals:     By definition, we have 
\begin{equation*}
\Tr[A_{g,\lambda,x}^p]
=\left (  \frac{\lambda }{1-x}    \right)^p       \int\frac{{\rm d}^{2p}   {\vec \alpha}}{\pi^p} ~   \prod_{j=1}^p   \left(   
e^{-\left(\lambda+1-\frac{1}{x}\right)|\alpha_j|^2} ~ \<  g  \alpha_j  |   g \alpha_{(j+1)\mod p}  \>  ~ \left  \langle \frac{\bar \alpha_j}{  \sqrt x}   \right  |  \left. \frac{\bar \alpha_{(j+1) \mod p}}{\sqrt{x}}\right\>\right),
\end{equation*}
where $\vec \alpha$  is the complex vector $\vec \alpha  : =  (\alpha_1,\dots, \alpha_p)^T   \in \mathbb C^p$ and ${\rm d}^{2p}\vec \alpha:=  \prod_{j=1}^p   {\rm d}^2  \alpha_j $.
Now, using the relation  $  \<  \alpha  |  \beta\>  =  e^{ \frac{- |   \beta |^2 -  | \alpha   |^2     + 2 \bar \alpha \beta  } 2 }  $
 we obtain  
\begin{align*}
\Tr[A_{g,\lambda,x}^p]
&=\left (  \frac{\lambda }{1-x}    \right)^p       \int\frac{{\rm d}^{2p}   {\vec \alpha}}{\pi^p} ~     
e^{-\left(\lambda+1-\frac{1}{x}\right)\|\vec \alpha\|^2} ~   e^{  g^2    ( -  \| \vec \alpha\|^2    +   \vec \alpha^\dag   S   \vec \alpha   ) }   ~ e^{  \frac 1 x    ( -  \| \vec \alpha\|^2    +   \vec \alpha^T   S   \vec {\alpha^*})}  ,  
\end{align*}
 where $S$ is the shift matrix defined by $S_{jk}  :=   \delta_{k,( j+1)\mod p }  $.
 Elementary algebra then gives 
\begin{align}
\nonumber \Tr[A_{g,\lambda,x}^p]
&=\left (  \frac{\lambda }{1-x}    \right)^p       \int\frac{{\rm d}^{2p}   {\vec \alpha}}{\pi^p} ~     
e^{-\left(\lambda+1 + g^2\right)\|\vec \alpha\|^2} ~   e^{   g^2   \vec \alpha^\dag   S   \vec \alpha    }   ~ e^{  \frac 1 x     \vec \alpha^\dag   S^T   \vec {\alpha}}  \\
&=\left (  \frac{\lambda }{1-x}    \right)^p       \int\frac{{\rm d}^{2p}   {\vec \alpha}}{\pi^p} ~     
     e^{       \vec \alpha^\dag   \Gamma_p    \vec {\alpha}} \label{det}, 
\end{align}
where   
\begin{equation*}
\Gamma_p=\left(
\begin{array}{cccccc}
\lambda+1 +g^2 &-g^2&0&\cdots&0&-\frac{1}{x}
\\-\frac{1}{x}&\lambda+1  +  g^2 &- g^2&\cdots&0&0
\\0&-\frac{1}{x}&\lambda+1 + g^2 &\cdots&0&0
\\ \vdots&{}&{}&\ddots&{}&{}
\\0&0&0&\cdots&\lambda+1 + g^2&- g^2
\\-g^2&0&0&\cdots&-\frac{1}{x}&\lambda+1 +g^2
\end{array}
\right)  
\end{equation*}

Now, $\Gamma_p$ is a circulant matrix, and, therefore, can be unitarily diagonalized using the discrete Fourier transform.  Hence, the Gaussian integral in Eq. (\ref{det}) can be computed with a simple change of variables, giving 
\begin{align}\label{norm}
\Tr [A_{g,\lambda,x}^p]   =       \frac{\lambda^p}{(1-x)^p       \det  \Gamma_p  }.
\end{align}
Taking the $p$-th root and the limit $p \to \infty$ we finally obtain
\begin{align*}
\|A_{g,\lambda,x}\|_\infty  =     \frac{\lambda}{(1-x)   \lim_{p \to \infty}   ( \det  \Gamma_p )^{\frac 1p} }.
\end{align*}

Now, the eigenvalues a circulant matrix are easily found by Fourier transforming its entries \cite{circu}. In our specific case, the eigenvalues of  $\Gamma_p$  are $\gamma_{p,n}  =  a-  b \omega_p^n   -  c \omega_p^{-n} $, with $\omega_p: =  \exp ({2\pi i } / p)$ and  $n =  0,\dots, p-1$. Hence,  we have 
\begin{align*}
 \lim_{p  \to\infty}    \ln    \left(\det\Gamma_p\right)^{\frac 1p} &  =  \lim_{p\to \infty}  \frac 1p   \sum_{n=0}^{p-1}    \ln ( a-  b \omega_p^n  -  c \omega_p^{-n} )\\
    &  = \int_{0}^{2\pi}  \frac {  d  \theta}{2\pi}  \ln (a  -  b  e^{i \theta}  - c e^{-i\theta}).    
\end{align*}
For $x\ge 1/(\lambda+1)$ we can decompose  $a  -  b  e^{i \theta}  - c e^{-i\theta}  = b(e^{i\theta} - y_+ )(y_-e^{-i\theta}  -  1) $  with
$y_{\pm}=\frac{\lambda+g^2+1\pm\sqrt{(\lambda+g^2+1)^2-4{g^2}/{x}}}{2g^2}$, we finally obtain  
\begin{align*}
\lim_{p  \to\infty}    \ln    \left(\det A_{g,\lambda,x}^p\right)^{\frac 1p}    =&  \int_0^{2\pi}\frac{ d \theta}{2\pi}    \ln   [ b(y_+  - e^{i\theta} )]  +  \ln[1 -y_-e^{-i\theta} ]\\
=&\ln(by_+),
\end{align*}
which, inserted in Eq. (\ref{norm}) gives
\begin{equation}\label{calcolofatto}
\|A_{g,\lambda,x} \|_{\infty}   =  \frac{2\lambda}{(1-x)  (\lambda+g^2+1+\sqrt{(\lambda+g^2+1)^2-4 {g^2}/{x}})}.
\end{equation}

Finally, we separate the two cases   $\lambda>g-1$ and $\lambda  \le g-1$.   For $\lambda  >   g  -  1$, we choose $x =\frac{g}{\lambda+ g + (g-1)^2}$ and obtain $  \|  A_{g,\lambda,x}  \|_\infty   =  \frac{\lambda}{\lambda+(g-1)^2}$, matching  the lower bound provided by (trivial) two-mode squeezing [Eq. (\ref{twomode})]. 
For $\lambda\leq g-1$, we choose $x  = 1/(\lambda +1)$ and obtain $\| A_{g,\lambda,x}  \|_{\infty}  =  (\lambda+1)/g^2 $, again, matching the lower bound provided by two-mode squeezing [Eq. (\ref{twomode})].  \qed

\section{Proof of Eq. (7): fidelity of optimal probabilistic amplifiers}

In the special case of Gaussian prior $p_{\lambda }  (\alpha)  =  \lambda e^{-\lambda |\alpha|^2}$ and for coherent input states $\rho_
 \alpha  =  |\alpha\>\<\alpha|$, the average state $\tau$ is the thermal state  $  \sigma_x  =  (1-x)  \sum_{n=0}^{\infty}  x^n |n\>\<  n| $  for    $ x  =  1/(\lambda + 1)$.   
 Then, using Eq. (\ref{calcolofatto}) for  $ x  =  1/(\lambda + 1)$ we get  
 \begin{align} \label{probabilistic}
 F^{prob}_{g, \lambda}    \le    \frac{2(\lambda + 1)}{ \lambda+g^2+1+  |  \lambda+1 - g^2|},   
 \end{align} 
 giving the bound $  F^{prob}_{g, \lambda}   \le    (\lambda+1)/g^2  $ for $\lambda   \le  g^2-1$ and $ F^{prob}_{g, \lambda}   \le 1$ for $\lambda   >  g^2-1$.

\section{Proof of Theorem 4: Optimal Design of Probabilistic Amplifiers }

To prove the theorem, we  exhibit suitable quantum operations that reach the fidelity in  Eq. (\ref{probabilistic}). 

\emph{1) Case  $\lambda  >   g^2 - 1$. }      For the quantum operation $\map Q_N  (\rho) =  Q_N \rho Q^\dag_N$ with $Q_N   \propto  \sum_{n=0}^N    g^n  |n\>\<  n| $, the  fidelity is given by  
\begin{align*}
F_{g,\lambda, N}     & =  \frac {  \int  \frac{{\rm d}^2\alpha}\pi    p_\lambda (\alpha)    ~       |  \<   g \alpha  |   Q_N   |\alpha\>|^2 }  {  \int  \frac{{\rm d}^2\beta}\pi    p_\lambda (\beta)    ~        \<   \beta  |   Q_N^\dag Q_N   |\beta\> } \\
 & =  \frac {  \int  \frac{{\rm d}^2\alpha}\pi    e^{- (\lambda+ 1 - g^2)|\alpha^2| }    ~       |  \<   g \alpha  |   P_N   |g \alpha\>|^2 }  {  \int  \frac{{\rm d}^2\beta}\pi     e^{- (\lambda+ 1 - g^2)|\beta^2| }     ~        \<   g \beta  |  P_N   |g\beta\> } \qquad P_N  :  = \sum_{n=0}^N  |n\>\<n|\\
 & \ge  {  \int  \frac{{\rm d}^2\alpha}\pi      (\lambda+ 1 - g^2)e^{- (\lambda+ 1 - g^2)|\alpha^2| }    ~      [  1- 2   \<   g \alpha  | (I- P_N)   |g\alpha\>  ] } \\
 & =    {1-  2   {  \left( \frac  {g^2}  {\lambda + 1}\right)^{N+1}}},
\end{align*} 
which converges to 1 exponentially fast  as $N$ increases. 

{\em 2) Case $g-1 < \lambda  \le  g^2-1$.}  For the quantum operation $\map Q_N  (\rho) =  Q_N \rho Q^\dag_N$ with $Q_N   \propto  \sum_{n=0}^N   x^n  |n\>\<  n| $,     the  fidelity is given by  
\begin{align*}
F_{g,\lambda, N}     & =  \frac {  \int  \frac{{\rm d}^2\alpha}\pi    p_\lambda (\alpha)    ~       |  \<   g \alpha  |   Q_N   |\alpha\>|^2 }  {  \int  \frac{{\rm d}^2\beta}\pi    p_\lambda (\beta)    ~        \<   \beta  |   Q_N^\dag Q_N   |\beta\> } \\
 & =  \frac {  \int  \frac{{\rm d}^2\alpha}\pi    e^{- (\lambda+ 1 - x^2) |\alpha^2| }    ~       |  \<   g \alpha  |   P_N   |x  \alpha \>|^2 }  {  \int  \frac{{\rm d}^2\beta}\pi     e^{- (\lambda+ 1 -x^2) |\beta^2| }     ~        \left \<  x \beta      \right|  P_N  \left | x\beta    \right\> }   \qquad P_N  :  = \sum_{n=0}^N  |n\>\<n|\\
 & \ge  {  \int  \frac{{\rm d}^2\alpha}\pi       (\lambda+ 1 -x^2)      e^{-  (\lambda+ 1 -x^2)  |\alpha^2| }    ~      [   |  \<   g \alpha    | x  \alpha \>|^2  - 2  | \<   g \alpha  | (I- P_N)   |x \alpha \> | ] } \\
 & \ge   \frac   {\lambda +1-x^2} {\lambda + 1 - x^2 + (g-x)^2}-  2   \sqrt  {   \mathbb{E}   (   \<   g \alpha  | (I- P_N)   |g \alpha \>   )      \mathbb{E}   (   \<   x \alpha  | (I- P_N)   |x \alpha \>   )  },
\end{align*} 
where $\mathbb E   (f_\alpha)$ denotes the expectation value of $f_\alpha$ over the Gaussian distribution  $p_{\lambda + 1- x^2} (\alpha)  =   (\lambda+ 1 -x^2)      e^{-  (\lambda+ 1 -x^2)  |\alpha^2| }  $.  Now,  it is easy to obtain  
\begin{align*}
  \mathbb{E}   (   \<   g \alpha  | (I- P_N)   |g \alpha \>   )   & =  \left( \frac{  g^2}{  g^2  +  \lambda + 1- x^2  }\right)^{N+1}\\
  \mathbb{E}   (   \<   x \alpha  | (I- P_N)   |x \alpha \>   )   & =  \left( \frac{  x^2}{   \lambda + 1  }\right)^{N+1} .
\end{align*}
Hence, for $x^2  < \lambda + 1$  the fidelity converges to  $\frac   {\lambda +1-x^2} {\lambda + 1 - x^2 + (g-x)^2}$ exponentially fast as $N$ increases.  If $\lambda <g^2-1$, the condition $x^2  < \lambda + 1$   is satisfied by choosing $  x=  (\lambda + 1)/g$, which gives fidelity $  (\lambda + 1)/g^2$ in the limit $N\to \infty$.   If $\lambda  =  g^2 -1$,  the condition $x^2  < \lambda + 1$   is satisfied by choosing $  x=  g-\epsilon$, which gives fidelity   $1 -  O(\epsilon^2)$ in the limit $N \to \infty$.

{\em Case 3)   $\lambda   \le  g-1$.}    Already treated in the deterministic case:  a two-mode squeezer is optimal here.   
  
\section{Proof of theorem 5: Benchmark for quantum amplifiers}  

\Proof  It is immediate to check that a heterodyne measurement followed by re-preparation of the state  $\left|  \frac{   g \hat \alpha}{1+\lambda }   \right\>$, corresponding to  the measure-and-prepare channel 
\begin{align}\label{hetero}  
\widetilde {\map C}  (  \rho)    =   \int    \frac{{\rm d}^2  \hat \alpha}{\pi}    ~     \<  \hat \alpha |  \rho  |\hat \alpha\>  ~    \left|  \frac{   g \hat \alpha}{1+\lambda }   \right\>   \left \<  \frac{   g \hat \alpha}{1+\lambda }   \right |  
\end{align}  
achieves the fidelity $\widetilde F_{g,\lambda}  =   \frac{    \lambda + 1}{  g^2  +  \lambda + 1 }$.  

We now prove that no measure-and-prepare channel can do better, both in the deterministic and in the nondeterministic case. Let us start from the deterministic case. Here we use Eq. (\ref{uppermp}) and the fact that $\|A_{g, p,\sigma}\|_\times =  \|A^{T_2}_{g,p,\sigma}\|_{\times}$, $T_2$ denoting the transposition on the second Hilbert space.     For the Gaussian distribution $p_{\lambda} (\alpha)  = \lambda  e^{-\lambda |\alpha|^2} $, we choose $\sigma$ equal to $\tau$,  the average state of the source, given by $\tau  =(1-x)  \sum_{n=0}^{\infty}  x^n  |n\>\<  n| $,  $x=  1/(1+  \lambda)$.  
Denoting the corresponding operator by $A_{g,\lambda, \tau}$, we have  
\begin{align*}
  A^{T_2}_{g,\lambda,\tau}&=\left ( \frac{\lambda}{1-x}\right) \int\frac{d^2\alpha}{\pi} ~e^{-\left(\lambda+1-\frac{1}{x}\right) |\alpha|^2} ~  |g\alpha\rangle\langle g\alpha|
\otimes  \left |\frac{ \alpha}{\sqrt{x}} \right \rangle \left\langle\frac{ \alpha}{\sqrt{x}} \right |\\
  &=\left ( \frac{\lambda}{1-x}\right) \int\frac{d^2\alpha}{\pi} ~e^{-\left(\lambda+1- \frac 1x \right) |\alpha|^2} ~   V^\dag_{\theta} \left(  |\sqrt {g^2  +  x^{-1}}   \alpha\rangle\langle \sqrt {g^2  +  x^{-1}}   \alpha| \otimes   |0\>\<0|   \right) V_\theta,  \end{align*}  
where $V_\theta  =  e^{\theta ( ab^\dag -  a^\dag b) }$ is a beamsplitter operator with $\theta  =  \tan^{-1}  (  g \sqrt{x}  )$. 
The calculation is particularly easy for $x=1/(\lambda + 1)$, where we have
 \begin{align*}
  A^{T_2}_{g,\lambda,\sigma}&=  \frac{\lambda}{(1-x)  (g^2  +  x^{-1})}    \int \frac{d^2\alpha}{\pi}    ~   V^\dag_{\theta} \left(  |  \alpha\rangle\langle  \alpha| \otimes   |0\>\<0|   \right) V_\theta \\
  &  =  \left(\frac{    \lambda + 1}{  g^2  +  \lambda + 1 } \right) ~  V^\dag_\theta  (I  \otimes |0\>\<  0|  )  V_{\theta} .
   \end{align*}  
Now, we have $  \|  A^{T_2}_{g,\lambda,\sigma}  \|_\infty  =  \frac{    \lambda + 1}{  g^2  +  \lambda + 1 }  =  \< 0  |  \<  0|  A_{g,\lambda,\sigma}^{T_2}  |0\>  |0\>  $.  
 By definition of the cross norm and of the operator norm, this implies
 \begin{align*}
 \|  A_{g,\lambda,\sigma} \|_\times = \|A^{T_2}_{g,\lambda,\sigma}\|_\times  =      \|  A^{T_2}_{g,\lambda,\sigma}  \|_\infty  =  \frac{    \lambda + 1}{  g^2  +  \lambda + 1 } .
 \end{align*}   
Using Eq. (\ref{uppermp}) we conclude that every measure-and-prepare channel $\widetilde{\map C}$  has fidelity $\widetilde F_{g,\lambda}  \le  ( \lambda + 1)/(  g^2  +  \lambda + 1)$.  This proves that the heterodyne measure-and-prepare channel of Eq. (\ref{hetero}) is optimal among all measure-and-prepare channels.  
 
 It remains to prove that the heterodyne channel is optimal also among  the \emph{probabilistic} measure-and-prepare protocols, described by quantum operations of the form $\widetilde {\map Q}  (\rho)  =  \sum_{ j  \in  \mathsf Y}    \Tr[ P_j \rho]     ~\rho_j$, with $  P_j \ge 0  \forall j \in  \mathsf Y$ and $ \sum_{j \in \mathsf Y}  P_j \le I $.  
 In this  case, the fidelity is given by  
 \begin{align*}
F^{prob}_{g, p}     &=    \frac {  \int  \frac{{\rm d}^2\alpha}\pi    p  (\alpha)    ~         \<   g \alpha  |   \map Q   (\rho_\alpha)  |g \alpha\> }  { \Tr [  \map Q (\tau)]}
\end{align*}
where $\tau$ is the average state of the source. 
Using Eq. (\ref{ausiliaria})  we then get  $F^{prob}_{g, p}     \le     \Tr [  A_{g,p,\tau}    \Sigma]$, where $\Sigma$ is the separable quantum state $ \Sigma := \Phi_{\tau, \widetilde{\map Q}}  /\Tr[\Phi_{\tau, \widetilde{\map Q}}] $. Hence, we obtain the bound  $F^{prob}_{g, p}     \le     \|  A_{g,p,\tau} \|_\times$.    Now, in the case of a Gaussian distribution we already showed in the first part of the proof that   $   \|  A_{g,p,\tau} \|_\times   =   ( \lambda + 1)/(  g^2  +  \lambda + 1)$.   This proves that  nondeterministic measure-and-prepare protocols have to satisfy the bound  $ F^{prob}_{g, p}    \le   ( \lambda + 1)/(  g^2  +  \lambda + 1)$.  Hence, the heterodyne measure-and-prepare channel of Eq. (\ref{hetero}) is optimal also among non-deterministic protocols.  
   \qed

\section{Relation between the optimal quantum channel and the optimal measure-and-prepare channel for $\lambda \le g-1$}

Consider the measure-and-prepare channel $\widetilde {\map C}_r$  defined as  
$\widetilde {\map C}_r (\rho)  :=          \int_{\alpha \in  \mathbb C}  ~   \frac {{\rm d}^2  \alpha }\pi   ~       |  \<  \alpha  |  0\>|^2  ~ |    \alpha \cosh r   \>\<    \alpha  \cosh r  |  . $  
Like the channel $\map C_r$, the measure-and-prepare channel $\widetilde {\map C}$  satisfies the covariance property
\begin{align*} 
\map C_r ( D(\alpha)  \rho   D(\alpha)^\dag)  =  D( \alpha \cosh r )  \map C_r  (   \rho )  D^\dag(\alpha\cosh r) \qquad  \forall \alpha\in\mathbb C,
\end{align*}
for every trace-class operator $\rho\in\map T(\spc H)$.  Moreover, we have $  \widetilde{\map  C_r}  ( |0\>\<  0|)  =  (1-y) \sum_{n=0}^\infty   y^n  |n\>\<  n|  $, with $y = \frac{\cosh^2 r}{  \cosh^2  r  +1 }$.  Recalling Eq. (\ref{param}) we then have $  \widetilde {\map C}_r (|0\>\<  0|)  =     \map A_{\cosh r/  \sqrt{\cosh r^2  + 1}}  \map C_{r'} (|0\>\<0|)$, where $\map A_{\cosh r/  \sqrt{\cosh r^2  + 1}}$ is an attenuation channel with attenuation parameter ${\cosh r/  \sqrt{\cosh r^2  + 1}}$, and $r':  =   \tanh^{-1}  \sqrt{  \frac{\cosh^2 r}{  \cosh^2  r  +1 } } $.   
Using the covariance properties of $\widetilde {\map C}_r$, $\map C_r$ and $\map A_{\cosh r/  \sqrt{\cosh r^2  + 1}}$ we then obtain  
\begin{align*}
 \widetilde {\map C}_r (|\alpha\>\<  \alpha|)  =     \map A_{\eta}  \map C_{r'} (|\alpha\>\<\alpha|)   \qquad \forall \alpha  \in  \mathbb C,
\end{align*}
which in turn implies 
\begin{align*}
 \widetilde {\map C}_r =    \map A_{\cosh r/  \sqrt{\cosh r^2  + 1}}  \map C_{r'} .
 \end{align*}  
Since the optimal quantum and classical channels are given by   $\map C_{g,\lambda}  =      \map C_{ \cosh^{-1}  [g/(\lambda + 1)]} $  and $\widetilde {\map C}_{g,\lambda}  =      \widetilde{\map C}_{ \cosh^{-1}  [g/(\lambda + 1)]} $ we have proven the relation  
\begin{align*}
\widetilde {\map C}_{g,\lambda}  =   \map A_{g/ \sqrt{g^2+(\lambda+ 1)^2}}  \map C_{ \sqrt{  g^2  +( \lambda  +1)^2} ,\lambda}. 
\end{align*}

\section{Coping with infinite dimensions: the truncation argument}

Here we show how to extend the validity of theorems 1 and 2 to the case when the input and/or output Hilbert spaces are infinite dimensional, by showing that the expression for the fidelity given therein can be achieved by a suitable sequence of quantum channels.  The proof is based on a truncation argument, that works when the average input state $\tau  =  \sum_{x\in\mathsf X}   p_x     \rho_x$---diagonalized as $  \tau  = \sum_{n=1}^{\infty}  p_n |n\>\<  n|$---has eigenvalues that decay sufficiently fast, in the sense that $\sum_{n=1}^{\infty}    p_n  E_n  =  E< \infty $ for some increasing sequence  ($E_{n+1}  \ge E_n \ge 0, \forall n \in \mathbb N$) such that $\lim_{n\to\infty}E_n =  \infty$. This is the case in all relevant examples: for example, a thermal state $\tau  =  (1-x)  \sum_{n=0}^{\infty}   x^n  |n\>\<  n|$ satisfies the required condition with $E_n  =  n$. 

We illustrate the truncation argument for general quantum channels, as the use of the argument for measure-and-prepare channels is exactly the same.  

Let us show that there exists a sequence of quantum channels reaching the value $F^{det}    =  \inf_{\sigma >  0,\Tr[\sigma]  =1}  \|  A_\sigma \|_{\infty}$.   
To prove the achievability of the value $F^{det}$, for every finite $N$ we define the value
\begin{align*}
F_{N}^{det}  = \inf_{\rho_N  \ge 0  ,  \Supp (\rho_N)  = \Supp (P_N),\Tr[\rho_N]  =1}  \left\|   A_{\rho_N} \right\|_{\infty}  \qquad A_{\rho_N}  :  = (I_{out}  \otimes  \rho^{-\frac 12}_N) A   (I_{out}  \otimes  \rho^{-\frac 12}_N)             
\end{align*} 
where  $P_N  =  \sum_{n=1}^N  |n\>\< n|$ and $\rho_N^{-1/2}$ is the inverse of $\rho_N^{1/2}$ on its support.    It is easy to see that the value $F_N^{det}$  is a lower bound to the fidelity that can be achieved by  quantum channels of the form $\map C_N  (\rho)  =  \map C_N   ( P_N  \rho  P_N  )  +   \Tr[(I_{in} -  P_N)  \rho]     \rho_0$, where $\rho_0$ is a fixed state.  Hence, 
\begin{align*}
\lim_{N\to \infty}   F_N^{\det}   \le  F^{det}.  
\end{align*} 

We now show that, in fact,  $\lim_{N\to\infty  }  F_N^{det}  =  F^{det}$.      Let $\rho_N$ be a state   such that  $\Supp(\rho_N)  =  \Supp (P_N)$, and  and let $\sigma_N$  be the  state 
\begin{align*}
\sigma_N : = p_N    \rho_N   +   \chi^T_N  \qquad  \qquad   p_N : =  \sum_{n=1}^N  \frac{ p_n E_n} E, \quad   \chi_N  :  = \sum_{n=N+1}^{\infty}   \frac{p_n  E_n} E   |n\>\<  n| 
\end{align*}
With this definition, we have 
\begin{align*}
\|  A_{\sigma_N}  \|_{\infty}   &  =       \|    A^{\frac 12}(I_{out}  \otimes \sigma_N^{-1}) A^{\frac 12}  \|_{\infty}  \\   &  \le      \frac 1 {p_N}    \|    A^{\frac 12}(I_{out}  \otimes \rho_N^{-1}) A^{\frac 12}  \|_{\infty}   +   \|    A^{\frac 12}(I_{out}  \otimes \chi_N^{-1})^T A^{\frac 12}  \|_{\infty}      \\
& = \frac 1 {p_N}     \|  A_{\rho_N} \|_{\infty}  +   \left \|   \left (  I_{out} \otimes  \chi_N^{-\frac 12}   \right)^T   A \left (  I_{out} \otimes \chi_N^{-\frac 12}    \right)^T    \right\|_{\infty} .              
\end{align*}
 Observe that, by construction,  the second term vanishes in the limit $N \to\infty$.  
 Indeed, we have     
\begin{align*}
 \left \|   \left (  I_{out} \otimes  \chi_N^{-\frac 12}   \right)^T   A \left (  I_{out} \otimes \chi_N^{-\frac 12}    \right)^T    \right\|_{\infty}       &=  \left \|    \left (  I_{out} \otimes  \chi_N^{-\frac 12}   \right)^T   \left(   \sum_{x\in\mathsf X}  p_x  |\psi_x\>\<\psi_x|  \otimes \rho^T_x   \right)      \left (  I_{out} \otimes  \chi_N^{-\frac 12}   \right)^T      \right\|_{\infty}  \\
 &\le  \left \|   I_{out} \otimes      \left (  \chi_N^{-\frac 12}   \tau \chi^{-\frac 12}_N\right)^T       \right\|_{\infty}  \\  
   &=  \left\|  \sum_{n=N+1}^{\infty}   \frac{E}{E_n}     |n\>\<n|    \right\|_{\infty}  \\
   &=       \frac{E}{E_{N+1}}  . 
\end{align*}
Hence, for sufficiently large $N$ we have 
\begin{align}
\left |\|   A_{\sigma_N}  \|_{\infty}   -    \| A_{\rho_N}  \|_{\infty} \right|  < \epsilon.
\end{align}  
In particular, choosing the state $\rho_N$ to satisfy 
\begin{align*}
\|  A_{\rho_N}  \|_{\infty}  <   F^{det}_N   +  \epsilon
\end{align*}
we obtain
\begin{align*}
F^{det}  \le  \|  A_{\sigma_N}\|_{\infty}   <  F^{det}_N  +  2 \epsilon, \end{align*}
and, therefore $F^{det}    \le \lim_{N\to \infty}  F_N^{det}$.  Since by definition  $F_N^{det}  \le  F^{det}$, for every $N$, this implies $F^{det}    = \lim_{N\to \infty}  F_N^{det}$.  Hence, the value $F^{det}$ can be achieved by a suitable sequence of quantum channels.

\end{widetext}

\end{document}